\RequirePackage[2020-02-02]{latexrelease}
\documentclass[12pt, notitlepage,aps,preprintnumbers,superscriptaddress,nofootinbib,aps,prd,floatfix]{revtex4-2}
\usepackage{float}
\usepackage{physics,slashed}
\usepackage{amsfonts,amsmath,amssymb}
\usepackage{mathrsfs}
\usepackage{bm}

\usepackage{epsfig}
\usepackage{graphicx}   
\usepackage{subfigure}  
\usepackage{verbatim}   
\usepackage{color}      
\usepackage{hyperref}   
\usepackage{url}
\definecolor{nicered}{rgb}{0.5,0.,0.}
\definecolor{nicegreen}{rgb}{0.,0.5,0.}
\definecolor{niceblue}{rgb}{0.,0.,0.5}
\hypersetup{colorlinks,citecolor=nicegreen,linkcolor=nicered,urlcolor=niceblue}
\usepackage{array}
\usepackage{dcolumn}
\usepackage{multirow}
\usepackage{cprotect}
\usepackage{framed}
\usepackage{setspace}

\newcommand{\GeV}{\textrm{GeV}}

\newcommand{\bea}{\begin{equation}\begin{aligned}}
\newcommand{\eea}{\end{aligned}\end{equation}}

\begin{document}
\title{Heavy Neutral Lepton Searches at the Electron-Ion Collider:\\ 
A Snowmass Whitepaper}
\preprint{HRI-RECAPP-2022-004, PITT-PACC-2206, Snowmass 2021: EF07, NF02, NF03, RF04}
\author{Brian Batell}
\email{batell@pitt.edu}
\affiliation{Pittsburgh Particle Physics, Astrophysics, and Cosmology Center, Department of Physics and Astronomy, University of Pittsburgh, Pittsburgh, PA 15206, USA\looseness=-1}
\author{Tathagata Ghosh}
\email{tathagataghosh@hri.res.in}
\affiliation{
Regional Centre for Accelerator-based Particle Physics,
Harish-Chandra Research Institute,
A CI of Homi Bhabha National Institute,
Chhatnag Road, Jhusi, Prayagraj 211019, India\looseness=-1}
\author{Tao Han}
\email{than@pitt.edu}
\affiliation{Pittsburgh Particle Physics, Astrophysics, and Cosmology Center, Department of Physics and Astronomy, University of Pittsburgh, Pittsburgh, PA 15206, USA\looseness=-1}
\author{Keping Xie}
\email{xiekeping@pitt.edu}
\affiliation{Pittsburgh Particle Physics, Astrophysics, and Cosmology Center, Department of Physics and Astronomy, University of Pittsburgh, Pittsburgh, PA 15206, USA\looseness=-1}
\collaboration{$\nu$-Test Collaboration}

\date{\today}
\begin{abstract}
In this whitepaper, we consider the model of heavy neutral leptons (HNLs) as an example to explore the potential of new physics searches at the Electron-Ion Collider (EIC). We propose two broad categories of search strategies depending on the HNL lifetime: direct searches for the prompt decay of HNLs with a short lifetime and displaced vertex searches for long-lived ones.
After identifying the most promising signals and the corresponding backgrounds, we perform a detailed simulation to estimate the sensitivity of the EIC to HNLs, accounting for detector thresholds, resolutions, and geometric acceptance. 
We derive projections for the EIC reach to the HNL squared mixing angle 
as a function of the HNL mass under the electron flavor mixing dominance hypothesis. Our findings indicate that the EIC can provide comparable sensitivity to the existing constraints for the prompt searches, while the displaced vertex searches can cover substantial new ground for HNLs in the 1-10 GeV mass range. Our proposed strategies are generally applicable to other new physics scenarios as well and motivate additional phenomenological exploration and dedicated future searches at the EIC.
%
\end{abstract}

\singlespacing
\maketitle

{\bf Introduction.} 
The future Electron Ion Collider (EIC) at Brookhaven National Laboratory will investigate the structure of nucleons and nuclei with unprecedented precision, allowing new insights into several key science questions, including the dynamics underlying nucleon spin and mass, the determination of parton distributions of nucleons in both momentum and position space, the confinement of hadronic states and nuclear binding, the saturation of gluon densities at a high energy, etc.~\cite{Accardi:2012qut}. 
To achieve these goals, the EIC will collide highly polarized electron and proton beams over a wide range of center-of-mass energies with a large integrated luminosity, 100-1000 times higher than the existing HERA dataset. To access a wide range of partonic momentum fraction $x$ and momentum transfer $Q^2$ in deep inelastic scattering, the detector is required to have broad momentum coverage with excellent tracking resolution and particle identification efficiency~\cite{AbdulKhalek:2021gbh}. 

Besides opening up a new QCD frontier, the EIC also offers an excellent chance to study precision electroweak (EW) physics and probe physics beyond the Standard Model (BSM)~\cite{Kumar:2016mfi}. These opportunities are facilitated by the large integrated luminosity, clean experimental environment, and multi-purpose hermetic detector~\cite{EIC-detector-handbook,AbdulKhalek:2021gbh}.
For example, a precise measurement of the EW neutral current at the EIC can sensitively probe higher dimensional operators in SM effective field theory,  (SMEFT)~\cite{Kumar:2013yoa,Zhao:2016rfu,Boughezal:2020uwq} and constrain the new neutral gauge bosons, \emph{e.g.}, $Z'$. Given the intense electron beam, the EIC provides an excellent laboratory to search for lepton flavor violation~\cite{Gonderinger:2010yn}, including that induced by axion-like particles (ALPs)~\cite{Davoudiasl:2021mjy}. 
Additionally, a standalone dedicated search for ALPs also appears to be promising~\cite{Liu:2021lan}.
Considering the substantial investment in the experimental facility, it is of great interest to broaden the EIC's physics program to include BSM physics searches. Among the range of possibilities, new heavy neutral leptons (HNLs) provide a well motivated BSM physics scenario that may intersect with several outstanding questions in particle physics and cosmology, including neutrino masses~\cite{Minkowski:1977sc,Yanagida:1979as,GellMann:1980vs,Glashow:1979nm,Mohapatra:1979ia,Schechter:1980gr}, the matter-antimatter asymmetry~\cite{Asaka:2005pn,Akhmedov:1998qx}, dark matter~\cite{Bertoni:2014mva,Gonzalez-Macias:2016vxy,Escudero:2016ksa,Batell:2017rol,Batell:2017cmf,Schmaltz:2017oov}, etc. 
In this whitepaper, we present our results of searching for a heavy neutral lepton ($N$) via the charged current process 
\begin{equation}
    e^- p \to N X
    \label{eq:NX}
\end{equation}
where $X$ is the beam remnant. 
We assume that $N$ decays to SM particles. 
We show that the EIC operating in the $ep$ mode will provide a powerful probe of HNL models, covering currently unexplored regions of mass-mixing angle parameter space. 

{\bf Heavy Neutral Leptons.}
%
In the standard Type-I seesaw model, HNLs, $N$, are introduced as gauge singlet fermions. HNLs couple to the SM particles through Yukawa-like interactions, with the new physics Langrangian written as
\begin{equation}
-\mathcal{L}\supset y\bar{\ell}_Li\sigma_2\Phi^*N+\rm{h.c.},
\end{equation}
where $\ell=(\nu,e)^T$ is the SM leptonic doublets, and $\Phi=(H^+,H^0)$ is the SM Higgs doublet. 
After electroweak symmetry breaking (EWSB), the Higgs vacuum expectation value will result in a Dirac mass term as $m_D\bar{\nu}_LN+{\rm h.c.}$ where $m_D=yv/\sqrt{2}$. Together with the HNL Majorana mass terms, $M_NN^TCN/2$, the light neutrinos acquire a tiny mass as $m_D^Tm_D/M_N$. 
In the mass basis, HNLs mix with the SM neutrinos with a mixing matrix $U$. For simplicity, here we take a phenomenological approach, assuming a single $N$ state in the mass range of interest to the EIC, with mixing dominated by the electron-flavor neutrino $\nu_e$. 
In such a scenario, only two parameters are relevant to the HNL phenomenology at the EIC, which are the HNL mass $m_N$ and the single mixing angle $|U_e|$. 
We also note that in specific scenarios the mixing angles can be significantly larger than the one suggested by the naive Type-I seesaw relation~\cite{Mohapatra:1986aw,Mohapatra:1986bd,Bernabeu:1987gr,Malinsky:2005bi}, resulting in an enhanced production rate of HNLs at the EIC.

\begin{figure}
    \centering
    \includegraphics[width=0.8\textwidth]{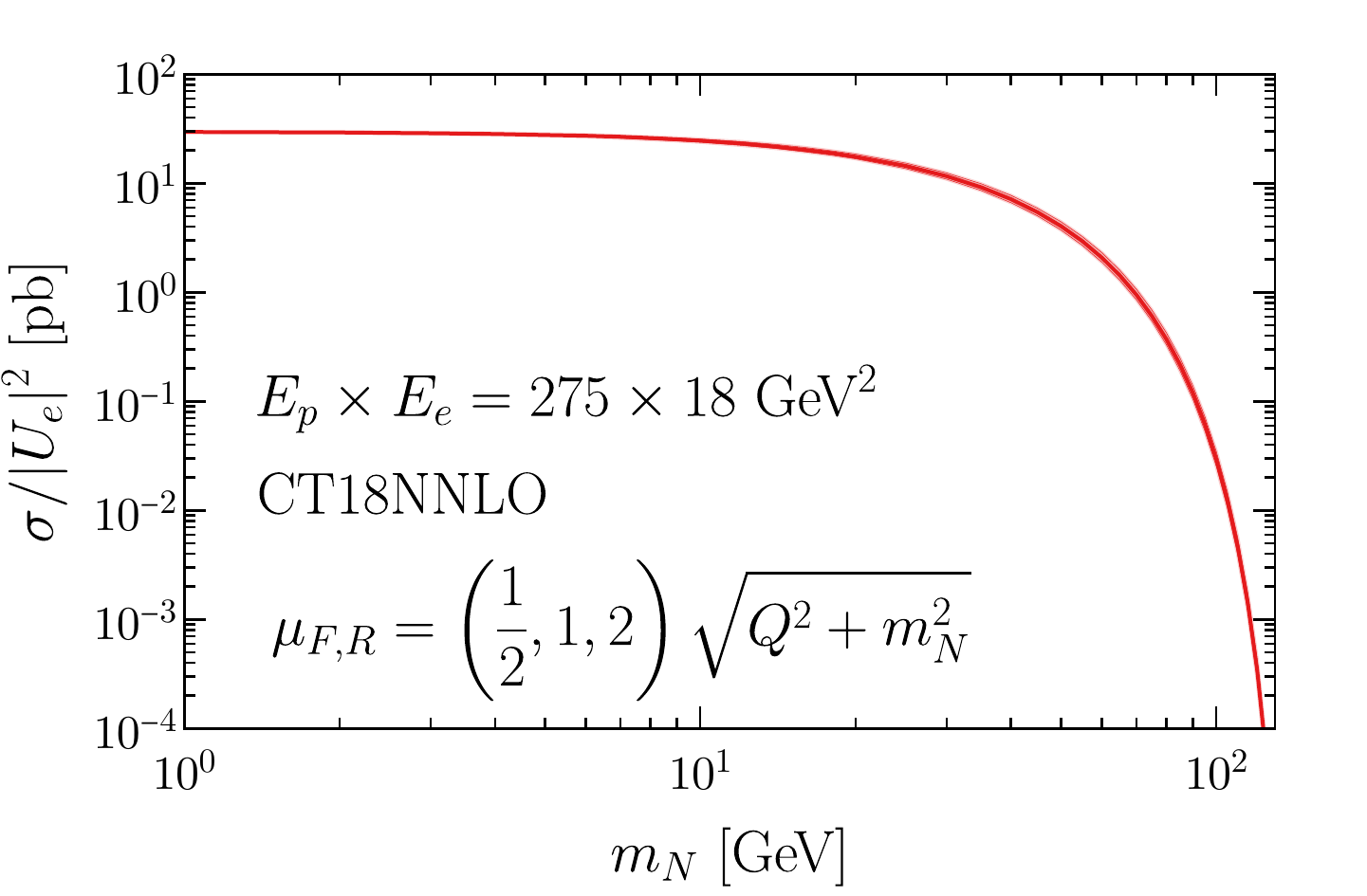}
    \caption{The HNL production cross section $\sigma$ [pb] divided by the mixing parameter $|U_e|^2$ at the Electron-Ion Collider with $E_p\times E_e=275\times18~\GeV^2$ beams versus the HNL mass $m_N$. The central prediction is based on CT18NNLO PDFs~\cite{Hou:2019efy} and renormalization/factorization scales as $\mu_{F,R}=\sqrt{Q^2+m_N^2}$, where $Q^2=-(p_e-p_N)^2$. The error band denotes the scale uncertainty by varying $\mu_{F,R}$ by a factor of 2, and the PDF uncertainty is negligible in this case.}
    \label{fig:xsec}
\end{figure}
The HNL couplings largely follow those of the SM neutrinos, which interact with electroweak gauge bosons, while the interaction strength is scaled by a factor $U$. In Fig.~\ref{fig:xsec}, we show the HNL production cross section for the process of Eq.~(\ref{eq:NX}) versus the mass $m_N$, 
using the optimally designed configuration of beam energies
\cite{AbdulKhalek:2021gbh}
\begin{equation}
    E_p=275~\GeV\ {\rm and}\  E_e=18~\GeV, 
    \label{eq:E}
\end{equation}
\emph{i.e.}, $\sqrt{s}=141~\GeV$.
The latest PDFs from the CTEQ-TEA group, CT18NNLO~\cite{Hou:2019efy}, are adopted as input to our predictions. The factorization and renormalization scales are chosen as
\begin{equation}
\mu_R=\mu_F=\sqrt{Q^2+m_N^2},
\end{equation}
where $Q^2=-(p_e-p_N)^2$ denotes the momentum transfer. The production cross section, which scales as $|U_e|^2$, can be as large as 30 (25) pb for $m_N=1~(10)~\GeV$ with $|U_e|^2=1$. The dominant theoretical uncertainty comes from the scale variation and is found to be a few percent for HNL masses in the GeV range and 20\% around $m_N\sim100~\GeV$, estimated by varying scales by a factor of 2, shown as error band in Fig.~\ref{fig:xsec}. 
In comparison, the PDF uncertainty is negligible. Considering the designed luminosity $\mathcal{L}=100$~fb$^{-1}$, EIC is expected to have sensitivity to squared of mixing angles of order $|U_e|^2\sim10^{-6}-10^{-4}$ if backgrounds can be brought under control.

\begin{figure}
    \centering
    \includegraphics[width=0.8\textwidth]{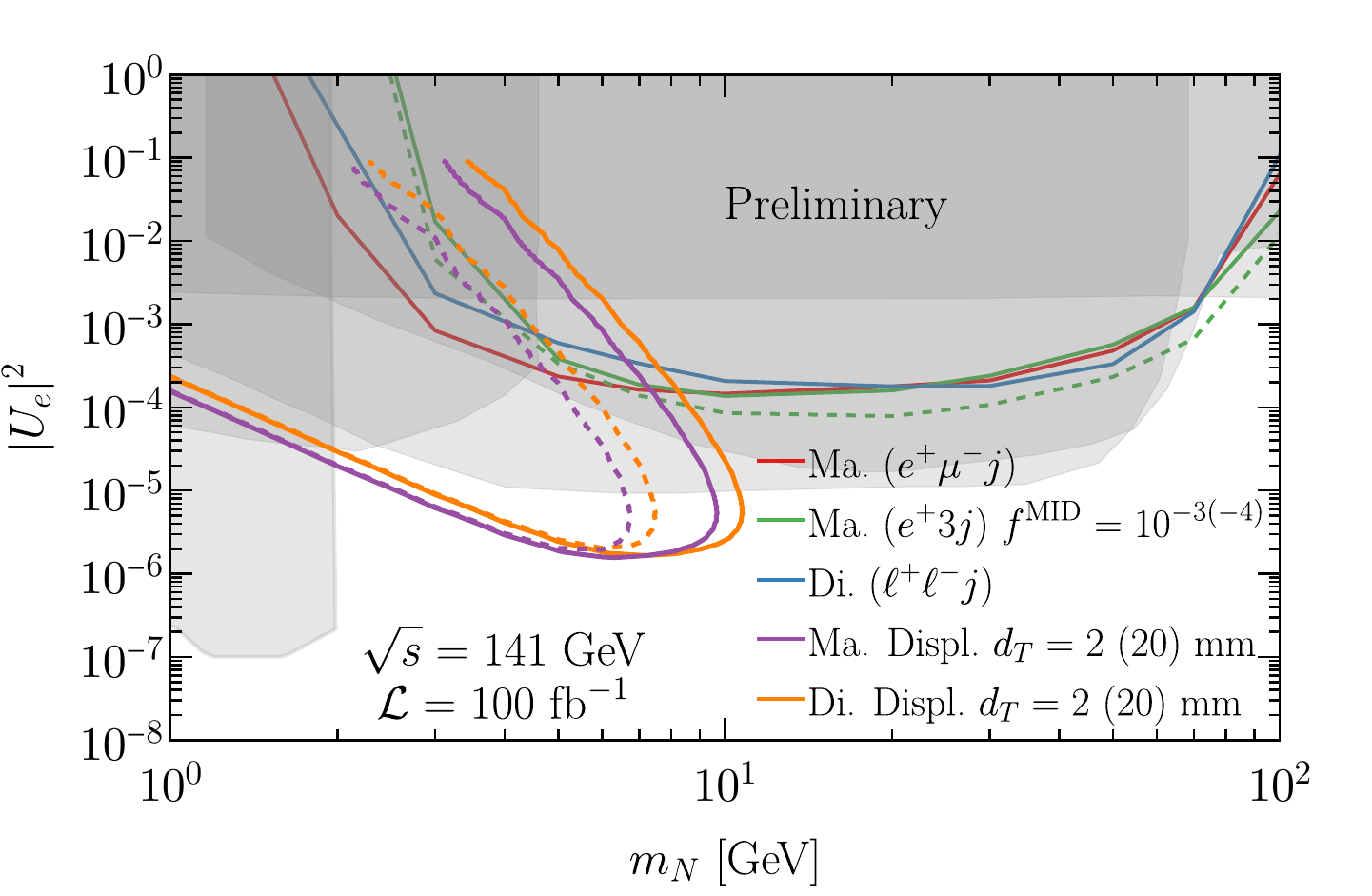}
    \caption{The EIC sensitivity to the HNL mixing squared angle  $|U_e|^2$ versus its mass $m_N$, in comparison with the existing searches~\cite{Beacham:2019nyx}. The solid (dashed) lines correspond to our assumption on electron charge mis-identification rates of $10^{-3}~(10^{-4})$ in the prompt search of $e^+3j$ channel, and transverse impact parameters as 2 (20) mm in displaced searches.}
    \label{fig:sense}
\end{figure}

{\bf Prompt Searches.}
HNLs will decay to the SM particles through weak interactions 
\begin{equation}
    N\to eW^*\ {\rm or}\  N\to\nu Z^* .
\end{equation}
In the standard Type-I scenario, both light neutrinos and HNLs are all Majorana fermions. A smoking gun signature for the Majorana HNLs would be lepton number violation by two units, corresponding to process $e^- p \to j(N\to e^+W^{-*})$. The virtual $W^{-*}$ will decay further into hadronic jets or into $\ell^-=e^-,\mu^-$ and a invisible neutrino. The di-electron channel $e^+e^-$ shares the same signature as $N\to\nu(Z^*\to e^+e^-)$. In comparison, $e^+\mu^-$ is a unique signature, which is almost background-free, and provides a great chance to probe the Majorana HNL at EIC. The main background comes from photon initiated production and subsequent decay $\gamma p\to (\tau^+\to e^+2\nu)(\tau^-\to\mu^-2\nu)+X$. With a full simulation of both signal and background, including detector effects, we obtain the EIC's sensitivity to $|U_e|^2$ down to $10^{-4}$ around the HNL mass $m_N\sim10~\GeV$, as shown the red line in Fig.~\ref{fig:sense}. The details of our simulation can be found in the forthcoming work~\cite{Batell:2022abc}.

Besides the $e^+\mu^-$ channel, the Majorana feature of lepton number violation can be also induced through $N\to e^+2j$ decay. The dominant background for this channel comes from the mis-identification (MID) of the electron charge, considering the large number of electrons $e^-$ entering detectors. Based on the current best MID rate of $0.1\%$~\cite{ATLAS:2019jvq,CMS:2015xaf} and a projected future one of $0.01\%$, we estimate the EIC sensitivity to HNL squared mixing angle $|U_e|^2$ at the same level or slightly better than the $e^+\mu^-$ channel, and shown as green solid (dashed) lines in Fig.~\ref{fig:sense}.

Besides collider searches, the neutrinoless double-beta decay $(0\nu2\beta)$ experiments have long been a powerful driver in the search for Majorana neutrinos through the search for lepton-number-violating processes~\cite{Dolinski:2019nrj,Bolton:2019pcu}.
In other scenarios, HNLs can be Dirac or quasi-Dirac, in which lepton number
violation is forbidden or highly suppressed~\cite{Mohapatra:1986aw,Mohapatra:1986bd,Bernabeu:1987gr,Malinsky:2005bi}.
Direct searches for Dirac HNLs have been carried out in meson rare decays, fixed-target experiments, and colliders~\cite{Bolton:2019pcu,Atre:2009rg,Cai:2017mow}. The EIC can also contribute to Dirac HNL searches. In this scenario, a relatively clean signature is provided by $N\to e^-(W^{+*}\to\ell^+\nu)$. 
Based our simulations~\cite{Batell:2022abc}, we obtain a similar EIC sensitivity to $|U_e|^2$ as in Majorana HNL searches, shown as the blue lines in Fig.~\ref{fig:sense}. In summary, prompt HNL searches at the EIC can be competitive with past constraints, particularly near the higher mass range of order $m_N \sim 100$ GeV.

{\bf Displaced searches.}
For smaller mixing angles, the HNLs become relatively long lived, leading to displaced decay signatures inside the EIC detector. 
A commonly used strategy for long lived particles involves the search for a displaced vertex. 
Alternatively, a displaced lepton with a large transverse impact parameter, $d_T$, can provide similar information, provided there is good tracking resolution. In consideration of the clean EIC environment, both approaches are expected to have low or negligible backgrounds. With the assumption of reasonable impact parameter measurement capabilities, such as $d_T=2~(20)$ mm, we have estimated the EIC sensitivity to $|U_e|^2$ for both Majorana and Dirac types of HNLs, shown as purple and orange solid (dashed) lines in Fig.~\ref{fig:sense}. 
We see that the EIC is capable of probing squared mixing angles down to $|U_e|^2\sim10^{-6}$ for masses $m_N\sim{\rm a~few}-10~\GeV$. The EIC reach extends significantly beyond existing constraints from LEP, potentially up to one magnitude in this region~\cite{DELPHI:1996qcc}.

{\bf Conclusions and prospects.}
In this whitepaper, we have investigated the potential of the EIC to directly search for HNLs via the SM weak interactions over the mass  range $m_N\in[1,100]$ GeV.
We have identified several promising signal channels and their associated SM backgrounds. Furthermore, we have carried out detailed simulations incorporating detector effects and designed kinematical and topological selection cuts to separate the HNL signal events from backgrounds. Based on these analyses, we have estimated the EIC sensitivities to the HNL squared mixing angle $|U_e|^2$ over the kinematically accessible mass range, with results summarized in Fig.~\ref{fig:sense}. 
We have proposed two broad classes of search strategies sensitive to both prompt and displaced HNL decays, which probe different regions of parameter space in the $(m_N,|U_e|^2)$ plane. For prompt decays, our strategy can provide comparable sensitivity to existing bounds from previous experiments~\cite{Beacham:2019nyx}. For the displaced vertex search, the EIC is expected to probe squared mixing angles down to $|U_e|^2\sim10^{-6}$, which exceeds the existing constraints from LEP by up to one magnitude in the several GeV mass region~\cite{DELPHI:1996qcc}. 

We also stress that our search strategies may be applicable to other new physics models that involve final-states with charged leptons and jets, as well as long-lived particles. Our results for the well motivated HNL model serve as an example study to illustrate the potential of the EIC in BSM searches and also may provide general guidance for the future experimental considerations, including e.g., detector designs. 

A detailed study of HNL phenomenology at the EIC will be presented in our forthcoming work~\cite{Batell:2022abc}.

{\bf Acknowledgments.}
We thank Yulia Furletova for helpful guidance about the EIC design during the early stage of this work. The work of BB, TH, and KX is supported by the U.S. Department of Energy under grant No. DE-SC0007914 and by PITT PACC. KX is also supported by the U.S. National Science Foundation under Grants No. PHY-1820760. TG would like to acknowledge the support from DAE, India for the Regional Centre for Accelerator based Particle Physics (RECAPP), Harish Chandra Research Institute.

\bibliographystyle{utphys}
\bibliography{EIC-HNL}
\end{document}